\documentclass[aps,prb,twocolumn,superscriptaddress,showpacs]{revtex4}

\usepackage{graphicx}

\usepackage{bm}
\usepackage{amsmath}

\begin{document}

\title{First order 0/$\bm \pi$ quantum phase transition in the
Kondo regime of a superconducting carbon nanotube quantum dot}

\author{Romain Maurand}
\affiliation{Institut N\'eel, CNRS et Universit\'e Joseph Fourier, BP 166, F-38042 Grenoble Cedex 9, France}
\author{Tobias Meng}
\affiliation{Institut f\"ur Theoretische Physik, Universit\"at zu K\"oln,
Z\"ulpicher Str. 77, 50937 K\"oln, Germany}
\author{Edgar Bonet}
\affiliation{Institut N\'eel, CNRS et Universit\'e Joseph Fourier, BP 166, F-38042 Grenoble Cedex 9, France}
\author{Serge Florens}
\affiliation{Institut N\'eel, CNRS et Universit\'e Joseph Fourier, BP 166, F-38042 Grenoble Cedex 9, France}
\author{La\"{e}titia Marty}
\affiliation{Institut N\'eel, CNRS et Universit\'e Joseph Fourier, BP 166, F-38042 Grenoble Cedex 9, France}
\email[]{laetitia.marty@grenoble.cnrs.fr}
\author{Wolfgang Wernsdorfer}
\affiliation{Institut N\'eel, CNRS et Universit\'e Joseph Fourier, BP 166, F-38042 Grenoble Cedex 9, France}


\date{\today}

\begin{abstract}

We study a carbon nanotube quantum dot embedded into a SQUID loop
in order to investigate the competition of strong electron correlations 
with proximity effect.
Depending whether local pairing or local magnetism prevails, a superconducting
quantum dot will respectively exhibit positive or negative supercurrent, referred 
to as a 0 or $\pi$ Josephson junction. In the regime of strong Coulomb blockade,
the 0 to $\pi$ transition is typically controlled by a change in the discrete charge state
of the dot, from even to odd. 
In contrast, at larger tunneling amplitude the Kondo effect develops for
an odd charge (magnetic) dot in the normal state, and quenches magnetism.
In this situation, we find that a first order 0 to $\pi$ quantum phase transition 
can be triggered at fixed valence when superconductivity is brought in, due
to the competition of the superconducting gap and the Kondo temperature. The SQUID 
geometry together with the tunability of our device allows the exploration of the 
associated phase diagram predicted by recent theories. We also report on the 
observation of anharmonic behavior of the current-phase relation in the transition 
regime, that we associate with the two different accessible superconducting states. 
Our results ultimately reveal the spin singlet nature of the Kondo ground state, which 
is the key process in allowing the stability of the 0-phase far from the mixed valence 
regime.

\end{abstract}

\maketitle

\section{Introduction} 
Realizing a Josephson junction with a carbon nanotube as a weak link opened up
the way to a new class of nanoelectronic devices combining both quantum confinement at the
nanoscale and the Josephson effect \cite{franceschi_2010, winkelmann_2009,hofstetter_2009, herrmann_2009,kasumov_1999, buitelaar_2002, jarillo-herrero_2006,dam_2006, cleuziou_2006, jorgensen_2007,cleuziouCM_2006,cleuziou_2007, pallecchi_2008,rasmussen_2009,
eichler_2009,kanai_2010}. In these junctions the critical current could be first thought to be 
maximized when a discrete electronic level of the quantum dot comes into resonance 
with the Cooper pair condensate of the electrodes, thus allowing an electrostatic tuning of the
magnitude of the critical current~\cite{beenakker1992}. The Josephson effect in quantum dots is 
however more complex, because it is governed by the interplay of electronic pairing and strong Coulomb 
interaction on the dot~\cite{rozhkov1999,clerk2000,yoshioka_2000,vecino_2003,siano_2004,choi_2004,sellier_2005,
novotny_2005,bauer_2007,karrasch_2008,meng_2009}.
When superconductivity dominates, the superconductor wave function spreads over the dot, 
inducing a BCS-singlet ground state, {\it i.e.} a standard Josephson junction 
(dubbed the $0$-state in what follows)~\cite{beenakker1992}.
In the other extreme regime of large electron-electron interactions, the quantum
dot enters the Coulomb blockade domain, and its charge is locked to integer values, 
altering the superconducting state.
For an odd occupancy, quantum dots behave like a spin $S=1/2$ magnetic impurity that
competes with Cooper pair formation, and the ground state can become a magnetic 
doublet.
In this situation, dissipationless current mainly transits through a fourth order tunneling 
process reordering the spins of Cooper pairs, thus leading to a negative sign of
the supercurrent, which is referred to as the $\pi$-type Josephson junction \cite{glazman_1989, dam_2006,
cleuziou_2006}. These two antagonist superconducting states, associated with a sharp sign reversal of 
the dissipationless current at zero temperature, can hence allow a first order 
quantum phase transition by tuning the microscopic parameters in the quantum dot. 
In the case of very strong Coulomb blockade, the 0-$\pi$ transition is achieved by 
modifying the parity of the electronic charge on the dot (valence is easily changed using
electrostatic gates), so that the supercurrent sign reversal occurs at the edges of the Coulomb diamonds. 
A more intriguing regime occurs for intermediate Coulomb repulsion (associated
to moderately small values of the tunneling amplitude compared to the charging
energy), in which Kondo correlations take place: in the normal state, the 
magnetic impurity of the odd charge state is screened through spin-flip cotunneling processes 
\cite{kouwenhoven_2001}, providing a non-zero density of states at Fermi energy. This so-called 
Kondo resonance allows the Cooper pairs to flow normally in the superconducting state, and a 
$0-$type Josephson junction is therefore recovered \cite{clerk2000, choi_2004, sellier_2005, 
eichler_2009, kanai_2010}. 
Here we explore in detail how superconducting transport is affected by the presence of 
Kondo behavior and we finely tune the 0-$\pi$ quantum phase transition in this more complex regime by controlling the quantum dot microscopic parameters.

\section{Characterization of the nanoSQUID}

\subsection{Sample fabrication}
Here we investigate supercurrent reversal in a carbon nanotube Josephson 
junction using the nano-SQUID geometry, which implements two Josephson 
junctions in parallel built with a unique carbon nanotube~\cite{cleuziou_2006}. The
single-wall carbon nanotubes were obtained using laser ablation and then dispersed 
in a pure dichloroethane solution using low power ultrasounds. 
A degenerately doped silicon wafer with a 450nm layer of SiO$_2$ on top was 
used as backgate. A first optical lithography step provided alignment marks 
in order to locate the nanotubes by scanning electron microscopy. The 
superconducting loops and the sidegates were fabricated using aligned e-beam 
lithography, followed by e-beam evaporation of Pd/Al bilayer (with respective thickness 4nm/50nm).
All measurements were performed in a dilution refrigerator with a base temperature
of $T=35$mK, and the filtering stages were similar to the ones performed in 
Ref.~\onlinecite{cleuziou_2006}. 
Samples were current-biased, either for DC or lock-in measurements (with an AC
amplitude of 10pA), in order to measure directly the switching current or the 
differential resistance of the device.
The nano-SQUID switching currents $I_{\rm sw}$ were detected {\it via} a digital
filter which monitors the estimated variance of the average DC voltage,
see Appendix~\ref{detection} and Ref.~\onlinecite{liu_1995}.
\begin{figure}[htbp]
\centering
\includegraphics[width=1\columnwidth]{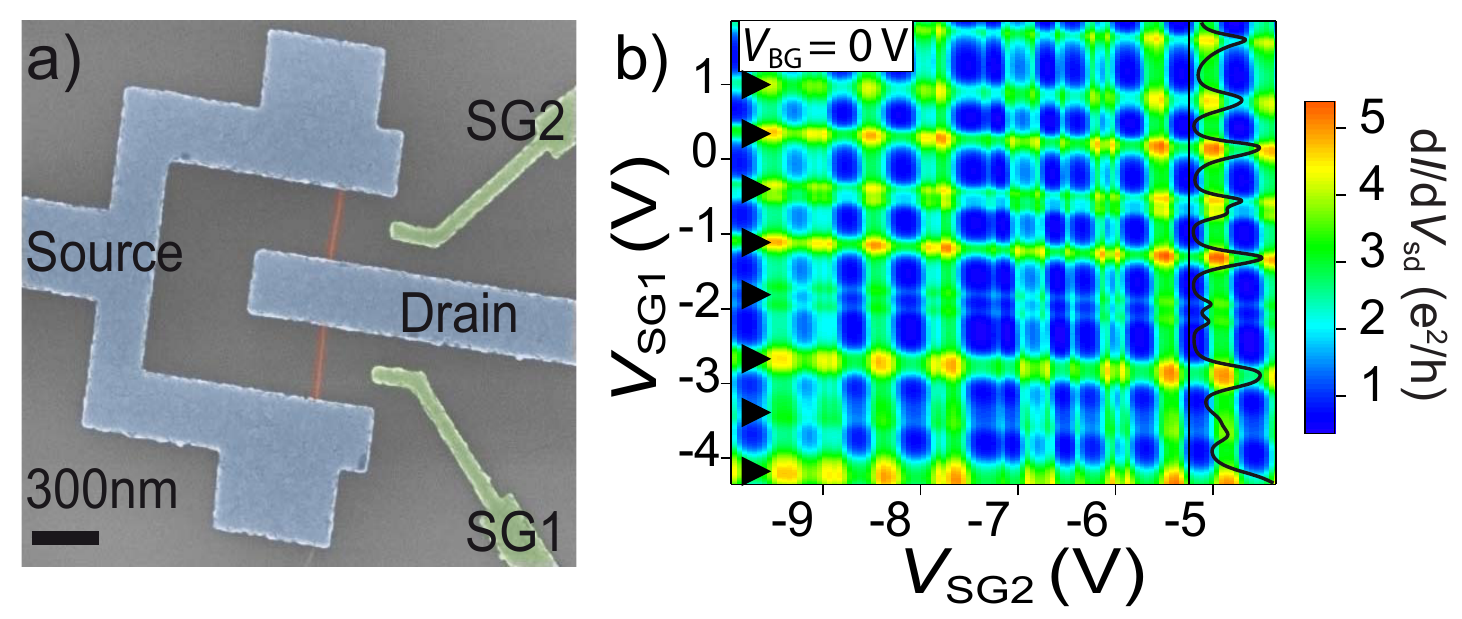}
\caption{ \textbf{Nano-SQUID characteristics. a)} SEM micrograph of the measured
nano-SQUID. The two sidegates (SG1,SG2) are colored in green, the nanotube
defining two quantum dots is shown in red, and the superconducting leads are in blue. 
\textbf{b)} Map of the normal state zero bias conductance $d I/d V_{sd}$ 
\textit{vs} the two sidegate voltages at magnetic field $B=75$mT,
temperature $T=35$mK and backgate votage $V_{\rm BG}=0$V.
Black triangles indicate the odd occupancy regions of the first quantum dot
(QD1). A line cut of the linear conductance at fixed $V_{SG2}=-5.25$V is also shown.}
\label{fig:squid}
\end{figure}
Fig.~\ref{fig:squid}a shows a SEM micrograph of the measured nanotube SQUID
with two $350$nm long nanotube Josephson junctions (JJ1 and JJ2).
Using the second quantum dot as a (possibly tunable) reference junction, a precise control over
the energy $\epsilon_0$ and linewidth $\Gamma$ of the first quantum dot is
accessible by tuning a pair of local sidegates and a backgate ($V_{\rm SG1}$, 
$V_{\rm SG2}$, $V_{\rm BG}$ respectively), see discussion below. 
Such a geometry allows us to directly measure the Josephson current of a single junction 
\textit{via} the magnetic field modulation of the SQUID switching current $I_{\rm sw}$, 
see Ref.~\onlinecite{cleuziou_2006}. Indeed, the critical current of an asymmetric SQUID 
with sinusoidal current-phase relation (taken here for simplicity) can be written as: 
\begin{equation}
I_c = \sqrt{\left(I_{c1}-I_{c2}\right)^2 
+4I_{c1}I_{c2}\left|\cos\left(\pi \frac{\phi}{\phi_0}
+\frac{\delta _1 + \delta_2}{2}\right)\right|^2}
\label{eq:squid}
\end{equation}where $\phi$ is the flux modulation of the SQUID, $\phi_0 = h/{2e}$ is the magnetic 
flux quantum, $(\delta_1,\delta_2)$ are the intrinsic phase shifts (0 or $\pi$) of the two Josephson 
junctions, and $(I_{c1},I_{c2})$ their respective critical currents. 
The critical current modulation is thus shifted by $\phi_0/2$ between the $0-0$
and the $\pi-0$ SQUID configuration.

\subsection{Normal state transport properties}
Fig.~\ref{fig:squid}b presents the nano-SQUID stability diagram $d I/d V_{sd}$ 
at zero bias \textit{vs} $V_{\rm SG1}$ and $V_{\rm SG2}$ in the normal state
(a perpendicular magnetic field of $B=75$mT is applied to suppress superconductivity), at a 
given backgate $V_\mathrm{BG}=0$V. This diagram 
resembles a weakly tilted checkerboard pattern, which is typical for two parallel 
uncoupled quantum dots in the Coulomb blockade regime, with a weak crosstalk of about 4\%. 
The line-cut at fixed $V_{\rm SG2}=-5.25$V emphasizes the regions of high and 
low differential conductance associated with the Kondo ridges and Coulomb
blockaded valleys respectively.  One can indeed distinguish easily between even and odd 
occupancies in each dot from the sequence of conducting and blocked regions:
dark blue pockets denote regimes where both dots are blocked (in an even-even
configuration of the double dot setup), green lines correspond to the situation of 
a single dot in the Kondo regime (see arrows) while the other remains blocked
(in an even-odd configuration), and red spots show the case where both dots undergo the Kondo 
effect (in an odd-odd configuration)~\cite{kouwenhoven_2001,cleuziou_2006}.\\
\begin{figure}[htbp]
\centering
\includegraphics[width=1.0\columnwidth]{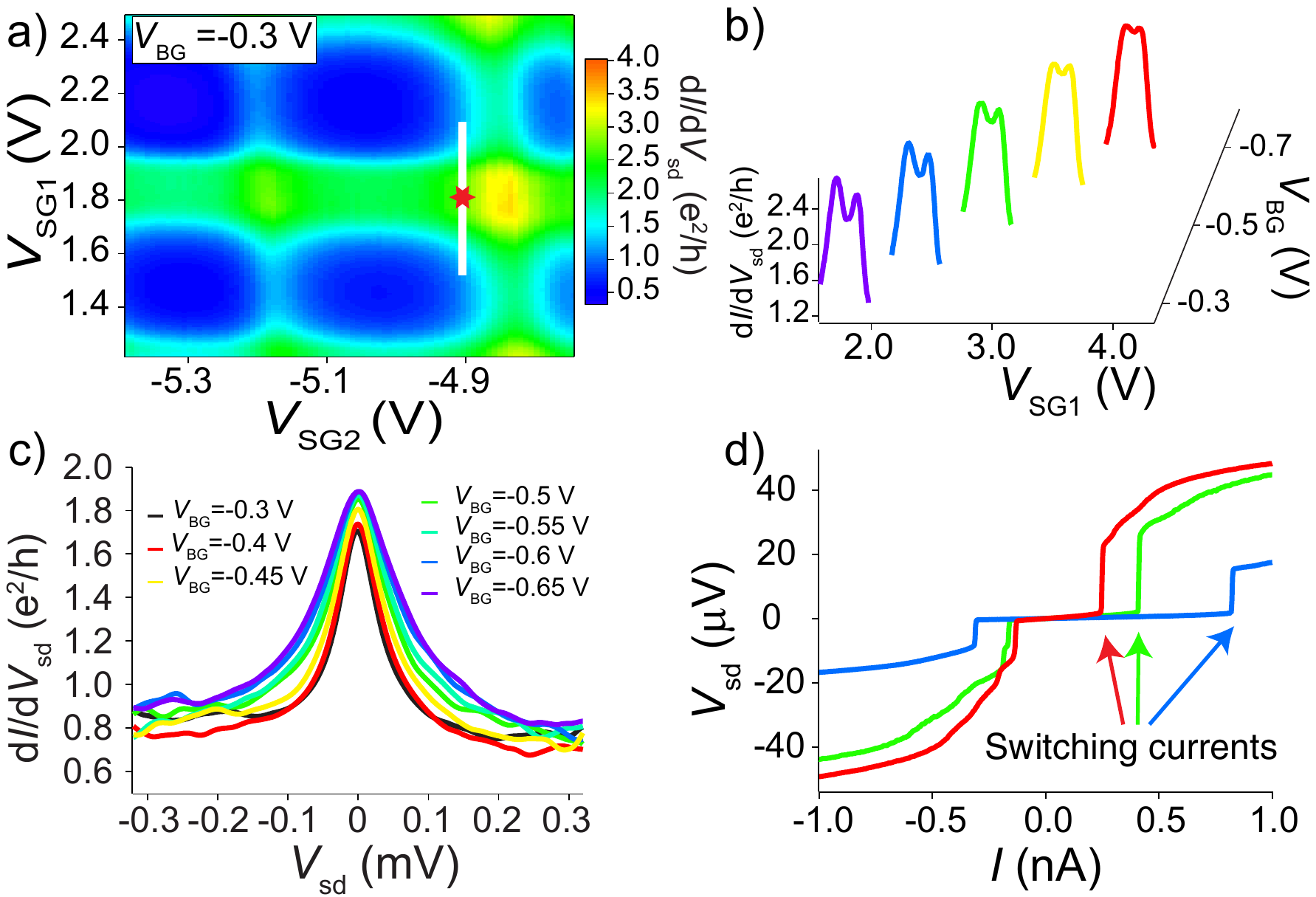}
\caption{\textbf{Kondo correlations and voltage/current characteristics. a)}
Zero bias conductance $d I/ d V_{sd}$ \textit{vs} the sidegate
voltages $V_{\rm SG1}$ and $V_{\rm SG2}$ in the normal state (under an applied 
magnetic field $B=75$mT) at $V_{\rm BG}=-0.3$V, where JJ1 presents a well established 
Kondo ridge for $1.7$V$<V_{\rm SG1}<1.95$V and JJ2 has an even occupancy for 
$-5.15$V$<V_{\rm SG2}<-4.85$V. 
\textbf{b)} Zero bias $d I/ d V_{sd}$ conductance \textit{vs} sidegate 
voltage $V_{\rm SG1}$ in the normal state along the white line in panel a) 
for five different backgate voltages $V_{\rm BG}$ between $-0.3$V and $-0.7$V. 
$V_{\rm SG2}$ was corrected for crosstalk in order to follow the white line, but
$V_{\rm SG1}$ was shown as measured.
\textbf{c)} Finite bias differential conductance $d I/ d V_{sd}$ \textit{vs} source-drain 
voltage $V_{\rm sd}$ in the normal state, taken in the middle of the Kondo ridge of JJ1 
(red star in panel a), corresponding to a level position $\epsilon_0=0$) for seven different 
backgate voltages $V_{\rm BG}$ between $-0.7$V and $-0.3$V. The width of the Kondo resonance
is modified by the backgate voltage while $\epsilon_0$ and $U$ are kept
constant, implying a variation of $\Gamma$. \textbf{d)} Typical superconducting 
voltage/current characteristics of the nano-SQUID for three arbitrary values 
of the gate voltages ($V_{\rm SG1}$,$V_{\rm SG2}$). The data are analyzed throughout the 
paper by recording the switching currents $I_{\rm sw}$ obtained from such 
voltage/current plot, see Appendix~\ref{detection} on the experimental technique 
that was used.}
\label{fig:Kondo1}
\end{figure}

An operating region at a different backgate voltage $V_{\rm BG}=-0.3$V is shown with greater 
detail on Fig.~\ref{fig:Kondo1}a. For $V_{\rm SG1}$ between $1.70$V and $1.95$V,
JJ1 has an odd occupancy associated with a differential conductance close to $2e^2/h$ 
due to a well-developed Kondo effect. Furthermore, JJ2 clearly has an even number 
of electrons for $V_{\rm SG2}$ between $-4.85$V and $-5.15$V, because of its
small contribution to transport in this range. In order to show the
influence of the backgate voltage $V_\mathrm{BG}$, we have plotted on Fig.~\ref{fig:Kondo1}b
the differential conductance \textit{vs} $V_{\rm SG1}$ for the odd occupancy
region of JJ1 corresponding to the white cut on Fig.~\ref{fig:Kondo1}a, taking five
different values of $V_{\rm BG}$ from $-0.3$V to $-0.7$V. By applying
$V_{\rm BG}$, the sidegates experience a capacitive crosstalk of $-21.5$\% and
$-17.4$\% for sidegate 1 and sidegate 2 respectively, as seen by the global
shifts of the conductance traces. The application of a backgate voltage thus modifies 
the occupation number on the dot, but also the tunnel linewidth $\Gamma$ 
\cite{grabert_1992, cleuziouCM_2006}. Indeed, by varying the backgate voltage and 
correcting the sidegates voltages for crosstalk, it is possible to keep
the local Coulomb repulsion $U$ and the level position $\epsilon_0$ on the
quantum dots relatively constant~\cite{kanai_2010}, while the hybridization $\Gamma$ of the first quantum dot (QD1) experiences sizeable variations up to about 20\%, as we discuss now.

\subsection{Tuning the hybridization with the gates}
\label{sec:Gamma}

On Fig.~\ref{fig:Kondo1}c, the Kondo resonances taken in the middle of the odd occupancy 
region of QD1 (see corresponding red star in Fig.~\ref{fig:Kondo1}a) are superimposed 
for different values of $V_{\rm BG}$. 
The hybridization $\Gamma$ can be extracted for different values of the backgate
voltage $V_{\rm BG}$ from the half-width at half-maximum
$V_K$ of the Kondo resonance in the finite bias conductance. 
In order to extract systematically $V_K$, we used a Lorentzian lineshape with 
fixed background corresponding to the QD2 contribution to transport and to a small elastic 
cotunneling component for QD1~\cite{goldhaber_1998}.
Qualitatively, we note the clear increase of $V_K$ that is achieved by shifting the 
backgate voltage to more negative values, which is related to the gate-induced enhancement 
of the hybridization $\Gamma$ reported above.
More precisely, in the scaling limit of the Kondo problem~\cite{haldane_1978},
a universal behavior of all physical observables is obtained as a function
of the Kondo scale, here expressed as a Kondo voltage $V_K$:
\begin{equation} 
V_K = \alpha\sqrt{\Gamma U}\exp\left[\frac{-\pi U}{8\Gamma}
\left(1-4\frac{\epsilon_{0}^{2}}{U^{2}}\right)\right] 
\label{eq:haldane2}
\end{equation} 
with $U$ the Coulomb repulsion on the dot, $\Gamma$ its total hybridization to
the leads, and $\epsilon_0$ its energy shift (taken by convention to zero in the
middle of the diamond).
\begin{figure}[htbp]
\centering
\includegraphics[width=0.8\columnwidth]{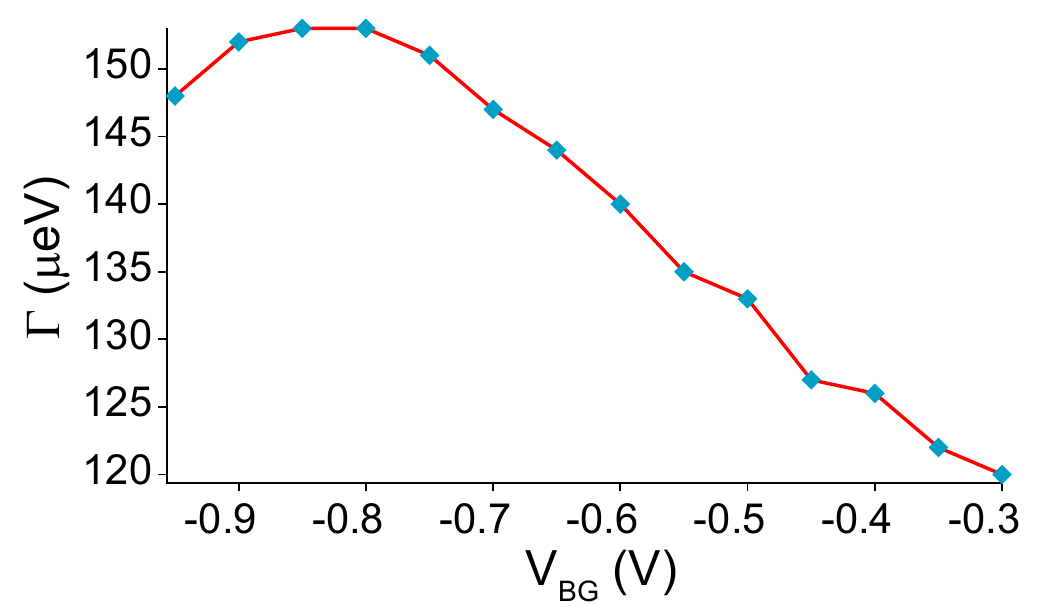}
\caption{{\bf Hybridization $\bf \Gamma$ \textit{vs} backgate voltage $\bf V_{\rm BG}$}. This
determination is performed in the middle of the Coulomb diamond of JJ1, see text for details.
A control with 20\% amplitude variations of $\Gamma$ is thus
achieved by tuning the backgate. For visibilty error bars are not indicated here, but are plotted later on Fig.~\ref{fig:Figure6}.}
\label{fig:gamma_vg}
\end{figure}
This expression applies in the limit $U\gg\Gamma$ (for $-U/2<\epsilon_0<U/2$), and 
contains a yet undetermined prefactor $\alpha$, which depends on the 
physical quantity under consideration. To obtain the value of the
charging energy $U$, we have considered the Coulomb stability diagram of JJ1, see
Appendix~\ref{U}, and extrapolated the diamond edges to large bias. Because of
the large linewidth of our strongly coupled nanostructure, the determination of
$U$ leads to moderate error bars, and we estimate $U= 0.80 \pm 0.05$ meV.
Now focusing on the differential conductance $\mathrm{d}I/\mathrm{d}V_{sd}$
from now on, a more precise definition of $\alpha$ is set by our choice of
$V_K$ as the half width at half maximum of the finite bias Kondo peak.
In a near equilibrium situation (corresponding to very asymmetric barrier to the left
and right leads) and in the regime $U\gtrsim6\Gamma$, we find that the unknown parameter is
given by $\alpha^\mathrm{eq}\simeq2.8$ from Numerical Renormalization Group (NRG) 
calculations~\cite{bulla2008}. However, in our experiment the conductance is tuned
to its maximum value $4e^2/h$ ({\it i.e.} $2e^2/h$ per dot), corresponding to equally 
balanced tunneling amplitudes from each leads (we note that values slightly above $4e^2/h$ can be
achieved in our double-dot device, which we attribute to small extra elastic
contributions from symmetry-broken orbital states of the carbon nanotube). 
In that case, decoherence of the Kondo anomaly is induced by the antagonist pinning of 
the Kondo resonance to the split Fermi levels of each lead, reducing the half-width $V_K$ of 
the finite-bias conductance peak compared to the equilibrium situation.
For the relevant regime $U\gtrsim6\Gamma$, one can estimate~\cite{fujii_2003} a
reduction by 50\% of the linewidth, so that we finally fix $\alpha=1.4$. Because
there is to date no fully controlled theory of the finite bias Kondo resonance,
we believe that the unprecise choice of $\alpha$ will introduce the largest error in
our determination of $\Gamma$, and hence of the phase boundary analyzed in
Sec.~\ref{sec:diag}.
The final backgate dependence of the hybridization $\Gamma$ is 
shown in Fig.~\ref{fig:gamma_vg} for the middle-point (particle-hole symmetric) 
of the Coulomb diamond of JJ1. The variations of $\Gamma$ with the backgate
$V_\mathrm{BG}$ are quite sizeable (up to 20\%), according to the exponential 
dependence of the Kondo scale~(\ref{eq:haldane2}) and constitute a central piece of the
analysis in the superconducting state, allowing to span a large part
of the phase diagram of the 0-$\pi$ transition. We also stress that changing the
local sidegates non-only allows to tune the energy levels in the dots, as is
clear in Fig.~\ref{fig:squid}b), but also modifies the hybridization $\Gamma$. 
The complete evolution of $\Gamma$ with backgate and sidegate voltages can be tracked 
by the analysis of the Kondo anomalies using Eq.~(\ref{eq:haldane2}), and leads to a greater 
range of variations (up to 50\%).

\section{Experimental study of the first order 0-${\bm \pi}$ transition}

Having characterized the normal state properties of our device, we now focus on 
the superconducting behavior of the nano-SQUID. 
Fig.~\ref{fig:Kondo1}d shows typical voltage-current characteristics obtained at three 
arbitrary gate voltages in the superconducting state. For all setpoints that we 
measured, the nano-SQUID shows an abrupt transition to the finite voltage branch 
indicating an underdamped device, with an hysteretic voltage-current characteristics 
\cite{cleuziou_2006,jarillo-herrero_2006}. The current at which this sharp jump 
occurs defines the switching current $I_{\rm sw}$, which can be precisely determined 
\textit{via} a digital filter \cite{liu_1995} calculating the maximum variance of 
the measured dc voltage, see Appendix~\ref{detection}. 
Switching currents of approximately $3$pA up to a few nA can thus be detected in a 
fully automated fashion.

\subsection{Comparison between valence induced and Kondo induced $\bm{0-\pi}$ transition}

\begin{figure}[htbp]
\centering
\includegraphics[width=1.0\columnwidth]{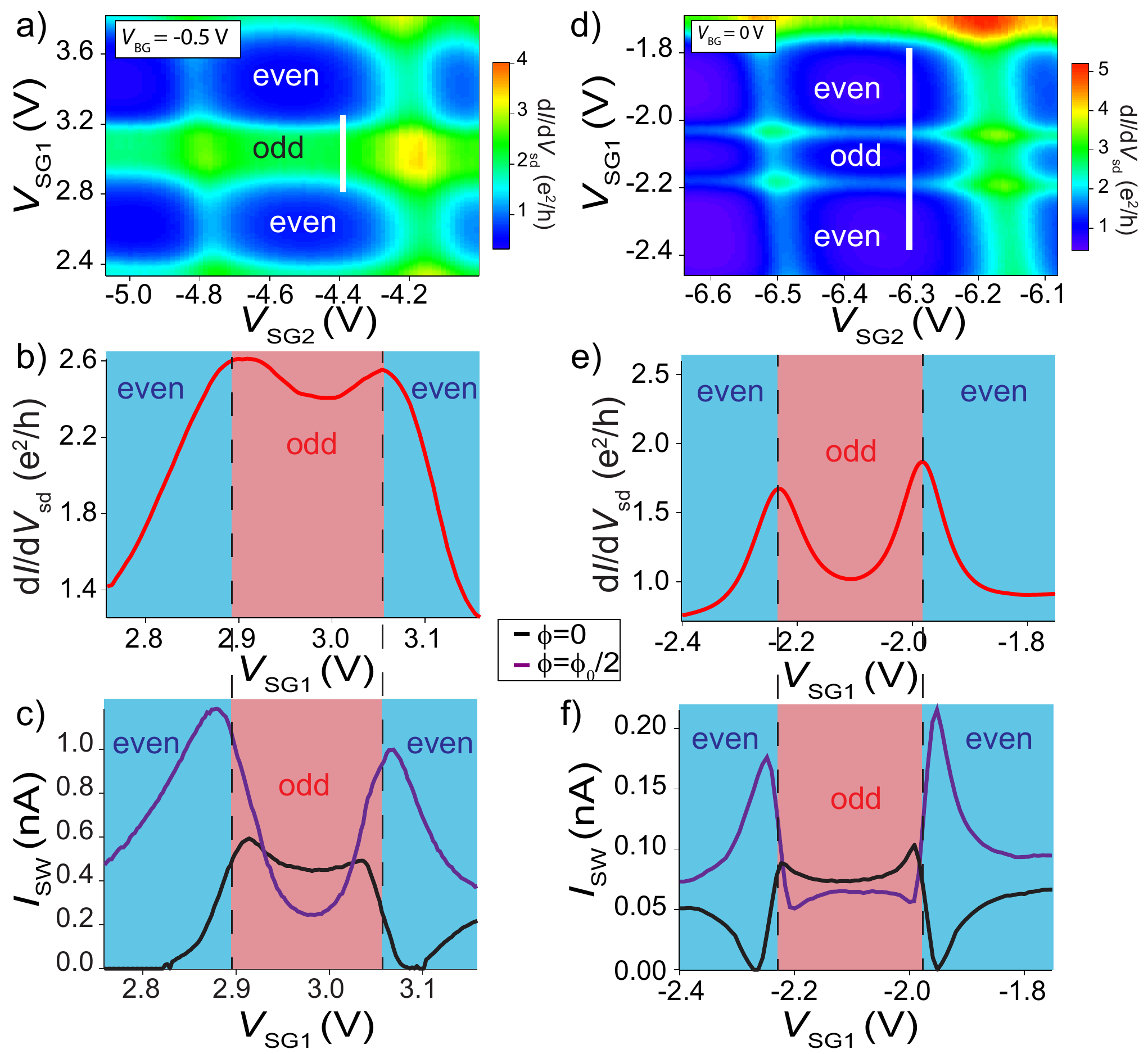}
\caption{\textbf{$\bm{0-\pi}$ transition in Kondo and Coulomb
blockaded odd-charge states.} Panels \textbf{a)}-\textbf{b)}-\textbf{c)} respectively show 
the stability diagram, the normal state conductance along the white line in a), and 
the supercurrent, in the case of a well-developped Kondo effect.
Panels \textbf{d)}-\textbf{e)}-\textbf{f)} respectively show 
the stability diagram, the normal state conductance along the white line in d), and the 
supercurrent, in the case when Coulomb blockade is not overcome by the Kondo effect. 
Supercurrent sign reversal, observed by the crossing of the two curves at $\phi=0$ and 
$\phi  = \phi_0/2$ respectively, penetrates in c) deep within the odd charge Coulomb diamond 
thanks to Kondo screening, in contrast to f), where the 0-$\pi$ crossing occurs
precisely when valence changes on the dot.}
\label{fig:CoulombKondo}
\end{figure}

Here we focus on a comparison of the 0-$\pi$ transition behavior in two
different correlation regimes, achieved in two distinct regions of the sidegate 
checkerboard diagram. Fig.~\ref{fig:CoulombKondo}a and
Fig.~\ref{fig:CoulombKondo}d show both operating regions:
a) corresponds to the situation already studied in the normal state, where fully
developed Kondo correlations take place for the odd region of JJ1, while d) reveals 
an odd charge state of JJ1 where Kondo correlations do not arise (Kondo temperature
smaller than the base temperature of the cryostat). 
These distinct physical regime are similarly witnessed on panels b) and e),
which show the normal state conductance trace along the white line in the 
conductance maps. 
In c), Coulomb blockade is fully overcome by the Kondo effect in the odd region
of JJ1, while in e) Coulomb blockade is robust throughout the entire gate range.
The most interesting comparison between the two regimes occurs in the
superconducting state, see panels c) and f). In panel c),
a supercurrent reversal (indicated by the crossing of the two curves associated
to two different magnetic flux, see Eq.~\ref{eq:squid}), occurs {\it within} the 
odd charge state of JJ1, showing that the Kondo effect plays a crucial role 
in triggering the $0-\pi$ transition. 
In contrast, panel f) shows that the supercurrent changes sign {\it
concomitantly with} the increase of valence of the dot, in agreement with
expectations in the strong Coulomb blockade regime~\cite{glazman_1989}.

\subsection{Tuning the 0-${\bm \pi}$ transition with controlled changes in the Kondo temperature}
\label{sec:diag}

\begin{figure}[htbp]
\centering
\includegraphics[width=1.0\columnwidth]{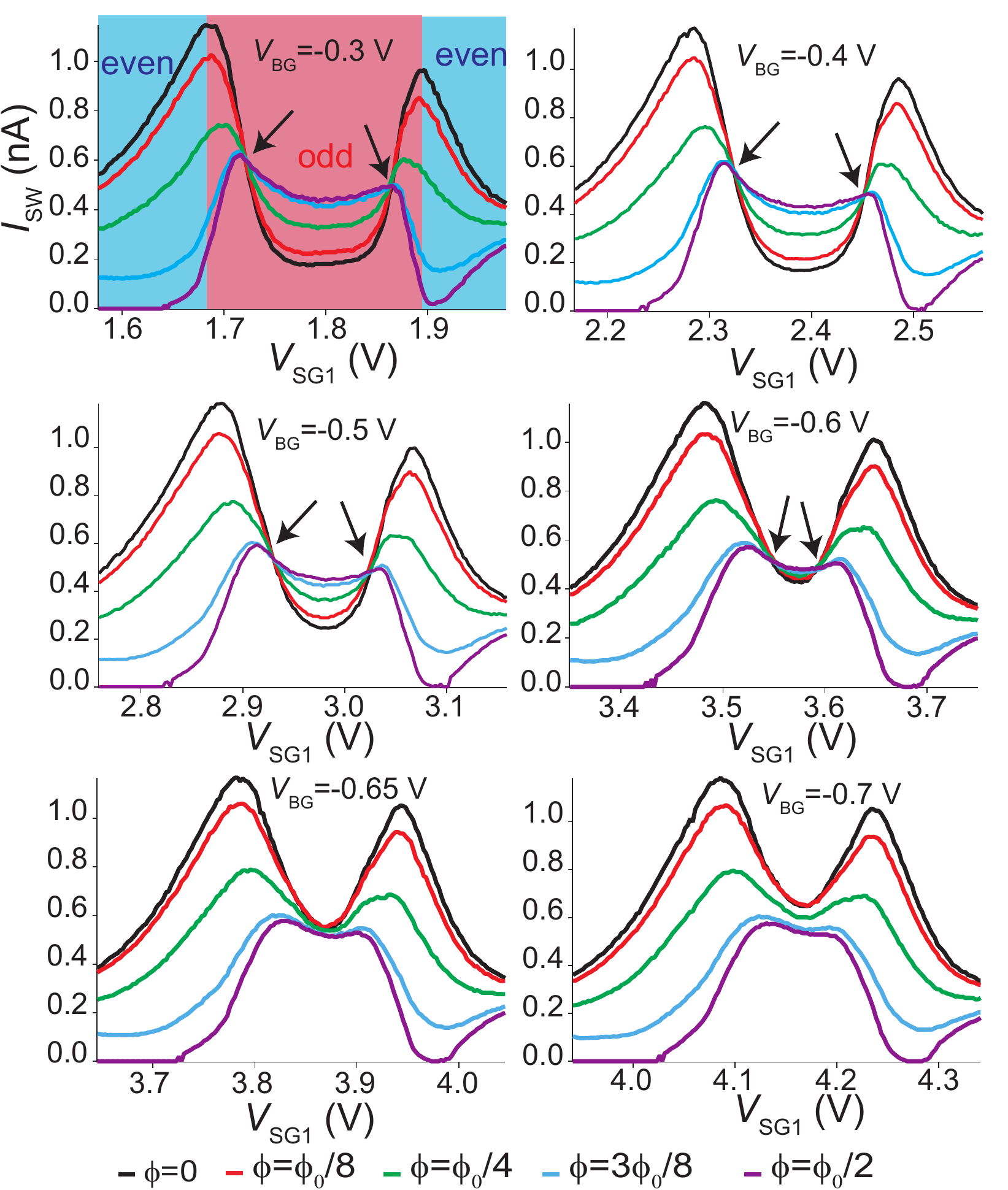}
\caption{\textbf{Switching current behavior of the nano-SQUID in the Kondo
regime}. The local gate voltage $V_{\rm SG1}$ 
dependence of the switching current $I_{\rm sw}$ is recorded for five different 
magnetic fields strengths (plotted as different colored lines). The various panels are
displayed by following the decrease of the backgate voltage $V_{\rm BG}$ from
-0.3V to -0.7V (the case $V_{\rm BG}=-0.5$V was 
previously shown in Fig.~\ref{fig:CoulombKondo}a,b,c). This decrease of the
backgate voltage allows to strengthen the Kondo effect, progressively shrinking the 
domain associated to $\pi$-junction behavior. Note that $V_{\rm SG2}$ was corrected for 
crosstalk in order to stay on the white cut shown in Fig.~\ref{fig:Kondo1}a, but the 
actually measured $V_{\rm SG1}$ is shown here. Arrows denote the transition region 
between $0$ and $\pi$-behavior, associated to the crossing point of the
switching current.}
\label{fig:mapchic}
\end{figure}

As shown in Sec.~\ref{sec:Gamma}, it is possible to tune the hybridization
$\Gamma$ with the gates, which we exploit to characterize more globally the 
0-$\pi$ transition phase boundary. 
Indeed, we saw previously that Kondo correlations in JJ1 are strengthened when 
$V_{\rm BG}$ goes from $-0.3$V to $-0.7$V in the operating region of Fig.~\ref{fig:Kondo1}a. 
In order to explore precisely the influence of Kondo correlations on the 0-$\pi$
transition, Fig.~\ref{fig:mapchic} presents six different plots of $I_{\rm SW}$ \textit{versus} 
$V_{\rm SG1}$ (along the white line on Fig.~\ref{fig:Kondo1}a) at different backgate voltages 
and magnetic fields.

\begin{figure}[tbp]
\centering
\includegraphics[width=1\columnwidth]{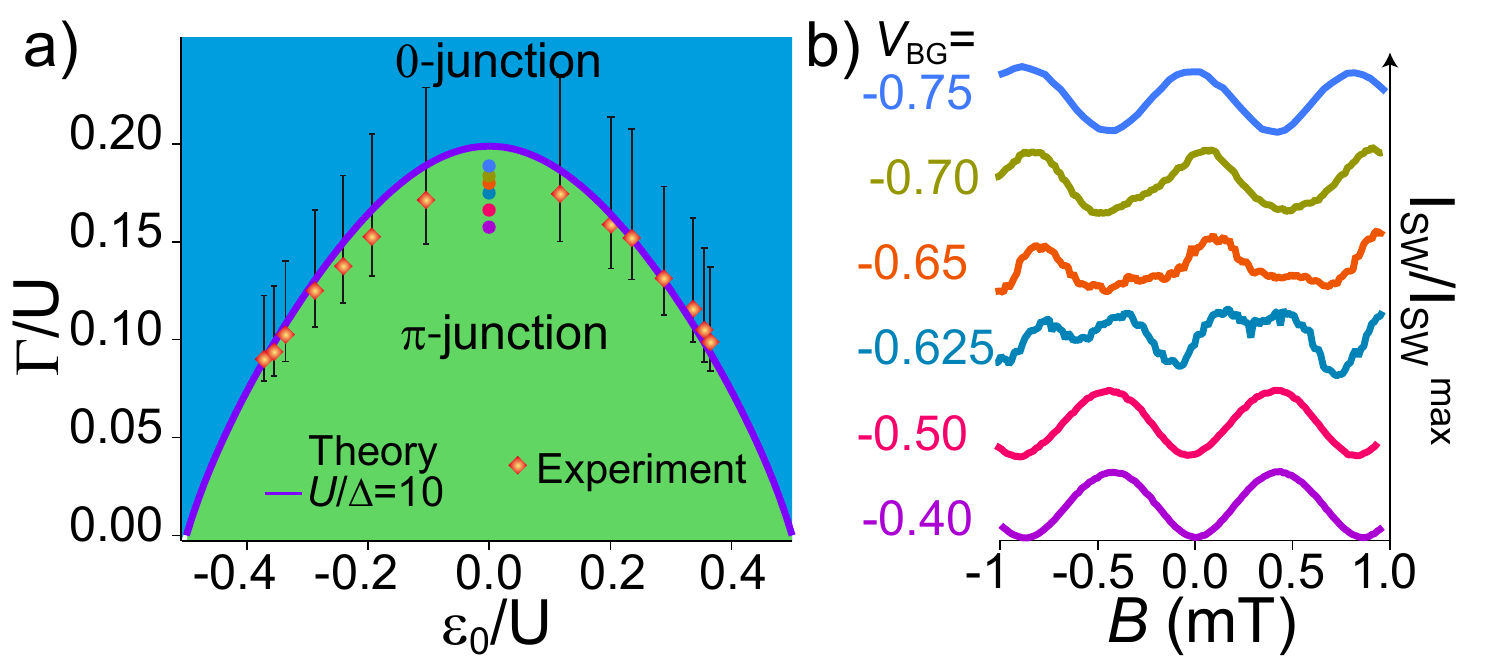}
\caption{\textbf{0-$\bm \pi$ phase transition diagram and nano-SQUID
modulations.} \textbf{a)} The experimentally determined phase boundary between 0 and $\pi$-junction 
behavior of JJ1 (orange squares) is given as a function of the dot energy $\epsilon_0/U$ 
and the level width $\Gamma/U$. The Coulomb repulsion $U \approx 0.8$meV was estimated 
from the finite bias spectroscopy of the Coulomb blockade diamond. Error bars indicate 
the uncertainty in the estimate of $\Gamma$ from the analysis of the Kondo resonances.
The purple line represents the theoretical phase diagram for $U/\Delta=10$ corresponding
to the experimental value of $\Delta \approx 80\mu$eV, see Appendix~\ref{delta}.
\textbf{b)} Magnetic field modulations of the nano-SQUID switching current 
taken in the middle of the odd-charge Kondo ridge (red star in
Fig.~\ref{fig:Kondo1}a) for different backgate voltages associated to the
fine mesh of dots at $\epsilon_0=0$ in panel a). For better comparison,
the switching current modulations are all normalised to the maximum current 
amplitude, which is strongly suppressed in the transition region.
A clear non-harmonic regime occurs near the $0$-$\pi$ phase boundary, where
bistable behavior (in phase) of the supercurrent may be attributed to the first 
order transition between the $0$ and $\pi$ states.}
\label{fig:Figure6}
\end{figure}

The traces exhibit two high switching current peaks corresponding to the Coulomb 
degeneracy points on the sides of the Kondo ridge in an odd valley of QD1. Recording 
such traces at different magnetic fields provides access to the flux modulation of 
the switching current in the nano-SQUID. Increasing the magnetic flux $\phi$ from 0 
to $\phi_{0}/2$ leads to a steady decrease of $I_{\rm sw}$ outside the odd 
occupancy region of JJ1, which corresponds to a standard 0-type behavior~\cite{cleuziou_2006,dam_2006}
in the Coulomb blockaded even valleys of QD1, see Eq.~\ref{eq:squid}.
The flux dependence of the switching current within the odd-charge Kondo domain
turns out to be more interesting, as we will analyze now. Clearly, the magnetic 
field behavior of $I_{\rm sw}$ is {\it reversed} deep inside the odd occupancy region of 
QD1, as the switching current is greater for $\phi=\phi_{0}/2$ than for $\phi=0$,
indicating a $\pi$-type Josephson behavior.
One can therefore identify precisely from the crossing of the switching current traces
at which sidegate voltage $V_{\rm SG1}$ (related to the dot energy) the behavior 
changes from 0 to $\pi$ type. This allows to define a $0$-$\pi$ phase
boundary for a given backgate voltage.
Now, by similarly examining the $I_{\rm sw}$ characteristics at different backgate voltages 
(allowing to tune the linewidth $\Gamma$), we note that decreasing the
backgate voltage ({\it i.e.} enhancing $\Gamma$) reduces the range for $\pi$ behavior, until 
the $\pi$ phase completely collapses below the critical $V_{\rm BG}=-0.65$V and
a $0$-junction is maintained all along the Kondo ridge. 
This physical behavior can be expected from the stronger Kondo screening at larger 
$\Gamma$ that tends to favor the $0$-state. 
From these measurements, we can unambiguously assign a $0$ or $\pi$ behavior to
the JJ1, as a function of both the level position $\epsilon_0$ and the
width $\Gamma$ of QD1, as determined previously from the analysis of the normal
state transport data. 
For all recorded transitions (corresponding to the black arrows in 
Fig.~\ref{fig:mapchic}), we have extracted the corresponding microscopic parameters 
$\Gamma$ and $\epsilon_0$, and plotted them on an experimental phase diagram
shown in Fig.~\ref{fig:Figure6}a. 
As a quantitative test of our analysis, we have displayed on Fig.~\ref{fig:Figure6}a 
the theoretical phase diagram obtained from a self-consistent description of Andreev
bound states \cite{meng_2009} for $U/\Delta \approx 10$ which corresponds to the experimentally
measured $U \simeq 0.8$meV and $\Delta \simeq 80\mu$eV. Error bars
are taking into account the uncertainty in the determination of $\Gamma$ from
the finite bias Kondo resonances, see Sec.~\ref{sec:Gamma}. The bell-shape of
the phase boundary together with the nearly quantitative comparison to theory
gives strength to the interpretation of the $0$-$\pi$ transition as a first order
phase transition associated to the crossing of the Andreev bound states at the Fermi level 
\cite{clerk2000,bauer_2007}. A key point here is that Kondo screening is
decisive to allow the existence of the 0-phase in the center of the odd charge
Coulomb diamond in our experimental conditions ($U\simeq6\Gamma$), see 
Fig.~\ref{fig:comparetheodiag} in Appendix~\ref{theory}.
While the 0-$\pi$ transition is always related to a simple Andreev level
crossing, this analysis of our data clearly demonstrates that it is the competition between 
the normal state Kondo temperature $T_K$ and the superconducting gap $\Delta$
which determines the precise location of the 0-$\pi$ phase boundary~\cite{bauer_2007,meng_2009}.

From theoretical expectations~\cite{rozhkov1999,karrasch_2008}, a second possible smoking gun 
for the first order 0-$\pi$ phase transition lies in the anharmonic behavior (in phase) of the 
Josephson junction in close vicinity to the $0$-$\pi$ phase boundary. 
This prediction motivates us to consider 
the field modulation of the switching current $I_{\rm sw}$ with fine changes of the
backgate voltage. This is plotted on Fig.~\ref{fig:Figure6}b, where the
quantum dot level was taken in the center of the odd Kondo valley (particle-hole 
symmetric point $\epsilon_0=0$, corresponding to the red star in Fig.~\ref{fig:Kondo1}a
and to the fine mesh of dots in Fig.~\ref{fig:Figure6}a).
For $V_{\rm BG}>-0.6$V modulations show that JJ1 is a $\pi$-junction because of
the $\phi_0/2$ shift from those of a normal SQUID. On the contrary for $V_{\rm
BG}<-0.7$V, modulations turn back to the standard behavior, indicating that JJ1 is a 
0-junction. However, for $-0.6>V_{\rm BG}>-0.7$ the nano-SQUID switching current modulations 
show strong anharmonicities. For this range of backgate voltage, the critical current of JJ1 is
very small, leading to a strongly asymmetric SQUID, thus JJ2 always switches at the same 
phase difference, implying that $I_{\rm sw}$ reflects directly the current-phase relation 
of JJ1 \cite{rocca_2007}. The observed non-harmonic signal could be interpreted
as a further indication for the bistable behavior of the junction associated with the
first order $0$-$\pi$ transition \cite{rozhkov1999,yoshioka_2000,clerk2000,choi_2004,karrasch_2008}. 

\section{Conclusion}
In conclusion, we have realized a nano-SQUID based on a superconducting carbon nanotube quantum
dots    that is fully tunable thanks a set of electrostatic gates, allowing a precise
control of the microscopic parameters of the device. This allowed us to determine 
an experimental phase diagram for the $0$-$\pi$ transition in the Kondo regime,
which turned in good agreement with theoretical calculations based on the competition between 
the Kondo temperature and the superconducting gap.
The observation of anharmonic behavior in the supercurrent-phase relation near the phase 
boundary is consistent with the first order nature of the $0$-$\pi$ transition
associated to the crossing of Andreev levels.
Fascinating prospects offered by the present work are the control and monitoring of 
the $0$-$\pi$ transition from supercurrent measurements, as performed here, with
{\it simultaneous} local spectroscopy of the Andreev spectrum on the quantum
dots in the spirit of the recent measurements of Pillet {\it et al.}~\cite{pillet_2010}. 
Such future developments of our experiment, which seems achievable by probing
the nano-SQUID with a scanning tunneling microscope, would bring strongly correlated 
superconducting nanostructures to a new level of control and understanding.

\begin{acknowledgments}
We thank T. Crozes, E. Eyraud, D. Lepoittevin, C. Hoarau, and R. Haettel for
their technical work, C. Thirion and R. Piquerel for their contributions in 
programming, and V. Bouchiat, H. Bouchiat, C. Balseiro, J.-P. Cleuziou, L. Bogani, 
S. Datta, D. Feinberg and T. Novotn\'y for useful discussions. The samples were 
fabricated in the NANOFAB facility of the N\'eel Institute. This work is partially 
financed by ANR-PNANO projects MolNanoSpin No. ANR-08-NANO-002, ERC Advanced Grant 
MolNanoSpin No. 226558, and STEP MolSpinQIP.
\end{acknowledgments}

\appendix

\section{Theoretical analysis of the $\bm{0-\pi}$ phase diagram}
\label{theory}

\subsection{Model for a superconducting quantum dot}

The standard Hamiltonian to describe a single superconducting quantum dot is given by the
superconducting Anderson model
\begin{equation}
\label{eq:Hamiltonian_complete_normal}
H = \sum_{i = L,R}{H_i^{}} + H_d + \sum_{i = L,R}{H_{T_i}} \text{ ,}
\end{equation}
where
\begin{eqnarray}
                \label{eq:H_lead} 
                H_i^{} &=& \sum_{\vec{k},\sigma}{\epsilon_{\vec{k}}^{} \,
c_{\vec{k},\sigma,i}^{\dagger} c^{\phantom{\dagger}}_{\vec{k},\sigma,i}} - \sum_{\vec{k}}{\left(
\Delta_{i}^{} \, c_{\vec{k},\uparrow,i}^{\dagger}
c_{-\vec{k},\downarrow,i}^{\dagger} + \mathrm{h.c.}\right) } \nonumber \\
                \label{eq:H_dot}
                H_d^{} &=& \sum_\sigma{(\epsilon_0^{}+U/2) \, d_{\sigma}^{\dagger}
d_{\sigma}^{}} + U n_{\uparrow}n_{\downarrow} \nonumber \\
                \label{eq:H_T_LD}
                H_{T_i}^{} &=& \sum_{\vec{k},\sigma}{\left( t_{}^{} \,
d_{\sigma}^{\dagger} c_{\vec{k},\sigma,i}^{} + \mathrm{h.c.}\right) } \nonumber
\text{ .}
\end{eqnarray}
In the above equations, $d_{\sigma}^{}$ is the annihilation operator of an
electron with spin $\sigma$ on the dot, $c_{\vec{k},\sigma,i}^{}$ that of an
electron with spin $\sigma$ and wave vector $\vec{k}$ in the lead $i= L,R$, and
$n_{\sigma} = d_{\sigma}^{\dagger}d_{\sigma}^{}$. The leads are described by standard 
s-wave BCS Hamiltonians $H_i$ with superconducting gaps $\Delta_{i}^{} = 
\Delta \, e^{i\varphi_i}$. The phase difference of the latter is noted 
$\varphi = \varphi_L - \varphi_R$. Furthermore, the leads are assumed
to have flat and symmetric conduction bands, i.e. the kinetic energy
$\epsilon_{\vec{k},i}^{}$ measured from the Fermi level ranges in $[-D,D]$ and
the density of states is $\rho_0=1/(2D)$. We assume $\vec{k}$-independent and
symmetric
tunneling amplitudes $t$ between the dot and both superconducting leads. The dot
is described by a single energy level $\epsilon_0$ submitted to the Coulomb interaction $U$
(in our convention $\epsilon_0$ vanishes at the center of the Coulomb diamond).

\subsection{Renormalized Andreev Bound states and phase diagram of the $0$-$\bm \pi$ transition}

In the superconducting state, the four atomic states of the quantum dot 
evolve onto renormalized Andreev bound states (ABS) that possibly live within the gap.
A quantitative description of this process was proposed in
Ref.~\onlinecite{meng_2009}, starting with bare values of the ABS splitting in
the limit of infinite gap:
\begin{eqnarray}
\label{eq:a0}
\delta E_-^0 &=& E_-^0 - E_{\sigma}^0 = \frac{U}{2} -
\sqrt{{\epsilon_0}^2+{\Gamma_{\varphi}}^2}\\
\label{eq:b0}
\delta E_+^0 &=& E_+^0 - E_{\sigma}^0 = \frac{U}{2} +
\sqrt{{\epsilon_0}^2+{\Gamma_{\varphi}}^2} \text{ .}
\end{eqnarray}
with
\begin{equation}
\label{eq:gamma_phi}
\Gamma_{\varphi} = \Gamma \frac{2}{\pi}
\arctan\left(\frac{D}{\Delta}\right)\cos\left(\frac{\varphi}{2}\right).
\end{equation}
In this simplified (and unrealistic) limit, the $0$/$\pi$ transition corresponds to the 
crossing of the $|-\rangle$ and $|\sigma\rangle$ states, which occurs for
$\delta E_-^0 = 0$, leading to a dome-like shape in the
($\epsilon_0/U$,$\Gamma/U$) plane. However, the phase boundary quantitatively
depends on the precise value of the superconducting gap $\Delta$, which must be
more realistically included in the calculation. This is done by calculating the corrections at
order $1/\Delta$ to the ABS positions~\cite{meng_2009}, followed by a
self-consistency loop that takes into account the leading logarithmic
singularities:
\begin{widetext}
\begin{eqnarray}
\label{eq:a}
\delta E_-(\Delta) & = & \delta E_-^0 -\frac{\Gamma}{\pi}\int_{0}^{D}{d\epsilon \,
\left[\frac{2}{E-\delta E_-(\Delta)}-\frac{1}{E+\delta E_+^0}-\frac{1}{E+\delta E_-^0}\right.}\nonumber\\
&+&\left.\frac{2\Delta}{E}uv\left|\cos\left(\frac{\varphi}{2}\right)\right|\left(\frac{2}{E-\delta
E_-(\Delta)}-\frac{1}{E+\delta E_+^0}+\frac{1}{E+\delta E_-^0}\right)\right]
+2|\Gamma_{\varphi}|uv 
\end{eqnarray}
and
\begin{eqnarray}
\label{eq:b}
\delta E_+(\Delta) & = & \delta E_+^0-\frac{\Gamma}{\pi}\int_{0}^{D}{d\epsilon \,
\left[\frac{2}{E-\delta E_+(\Delta)}-\frac{1}{E+\delta E_+^0}-\frac{1}{E+\delta E_-^0}\right.}\nonumber\\
&+&\left.\frac{2\Delta}{E}uv\left|\cos\left(\frac{\varphi}{2}\right)\right|\left(\frac{-2}{E-\delta
E_+(\Delta)} -\frac{1}{E+\delta E_+^0}+\frac{1}{E+\delta E_-^0}\right)\right]
-2|\Gamma_{\varphi}|uv \text{ ,}
\end{eqnarray}
\end{widetext}
with $E=\sqrt{\epsilon^2+\Delta^2}$, and $\delta E_-^0,\delta E_+^0$ have been defined in Eqs.
(\ref{eq:a0})-(\ref{eq:b0}). The numerical resolution of the self-consistent
equation~(\ref{eq:a}) provides an accurate determination of the phase boundary
under the condition $\delta E_-(\Delta)=0$, that we successfully compared to the
experimental data. Note that we corrected here a misprint in
Ref.~\onlinecite{meng_2009}, namely a factor 2 in front of the
second term within the integral in~(\ref{eq:a})-(\ref{eq:b}).

\begin{figure}[htbp]
\centering
\includegraphics[width=0.8\columnwidth]{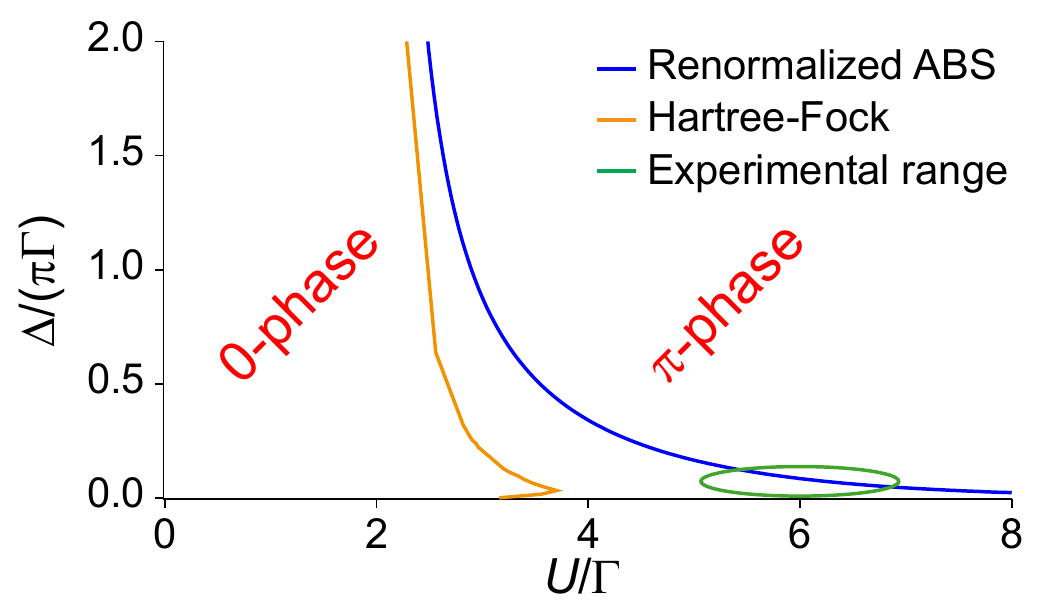}
\caption{\textbf{Theoretical phase diagram for the 0-${\bm \pi}$ transition at 
particle-hole symmetry.}
A comparison between the renormalized-ABS theory (which includes Kondo
correlations) and static Hartree-Fock theory is made. Computations were
done for the large bandwidth limit of the superconducting Anderson model
Eq.~(\ref{eq:Hamiltonian_complete_normal}), and the experimental range of
the operating region of Figs.~\ref{fig:mapchic} and \ref{fig:Figure6} was 
added for clarity. This comparison shows the key role of Kondo screening to allow
the existence of a 0-$\pi$ transition at intermediate correlations
($U>\pi\Gamma$).}
\label{fig:comparetheodiag}
\end{figure}
In order to stress the key role of the Kondo effect for the 0-$\pi$ transition
in our experimental conditions, we compare the phase diagram at particle-hole
symmetry ($\epsilon_0=0$) obtained from the renormalized-ABS
theory~\cite{meng_2009} and from {\it static} Hartree-Fock mean-field
theory~\cite{rozhkov1999}, see Fig.~\ref{fig:comparetheodiag}.
Because the renormalized-ABS approach includes the Kondo scale (at one-loop
order), it allows the extension of the 0-$\pi$ boundary for arbitrary large
values of $U/\Gamma$. In contrast, the static mean-field approach is unable
to restore a $0$-state for Coulomb interaction such that $U\gtrsim\pi\Gamma$,
and fails to reproduce our experimental observation of a supercurrent reversal
in the regime $U\simeq6\Gamma$. This comparison shows that the phase boundary
in our experiment is indeed associated to a competition between the normal state 
Kondo temperature and the superconducting gap, in agreement with theoretical
expectations~\cite{bauer_2007,meng_2009}.

\section{Determination of the microscopic parameters of the nanosquid}

\subsection{Charging energy}
\label{U}
\begin{figure}[htbp]
\centering
\includegraphics[width=1.0\columnwidth]{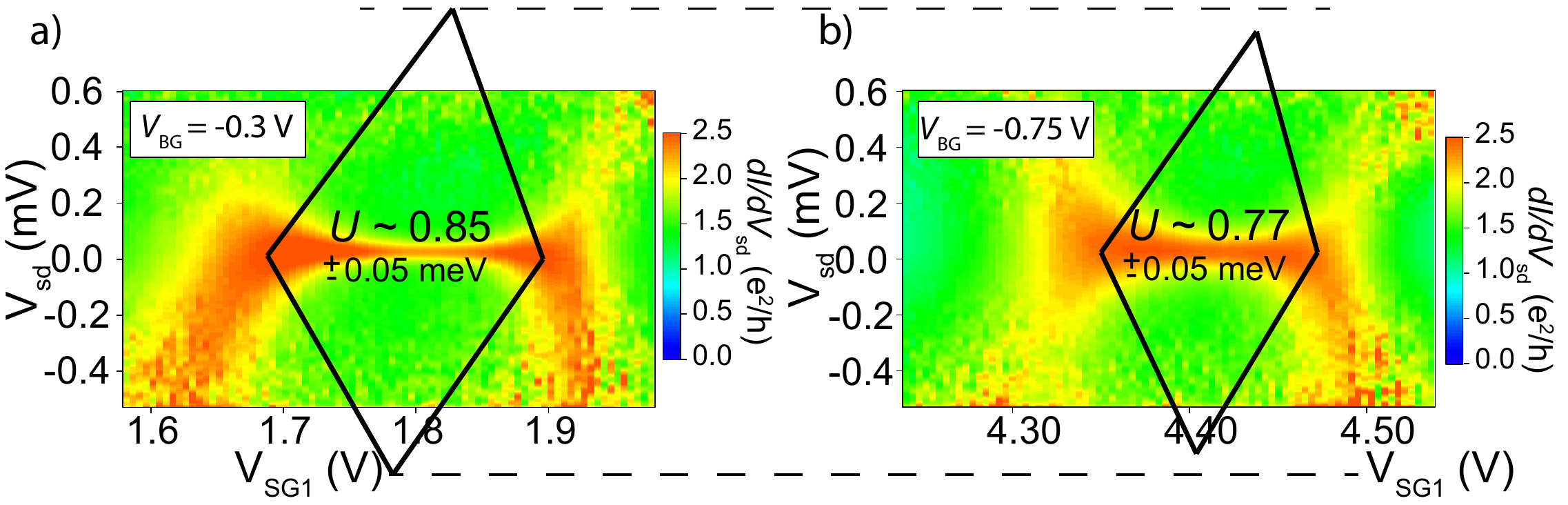}
\caption{{\bf Conductance map of JJ1 for two values of the backgate voltage.}
Extending the diamond edges to finite bias allows to extract the Coulomb repulsion 
of the dot $U=0.80\pm0.05$meV.}
\label{fig:U}
\end{figure}

Because the charging energy $U$ in a carbon nanotube quantum dot results
from the confinement between fixed contacts, one does not expect large
variations of $U$ for small detuning of the backgate. In order to determine
the experimental phase diagram for the $0-\pi$ transition, an estimate
of $U$ is required. This is obtained by considering the Coulomb stability
diagram of JJ1 for two different values of the backgate, see Fig.~\ref{fig:U},
and extrapolating the diamond edges to large bias.
Because of the large linewidth of our strongly coupled nanostructure, the
determination of $U$ also contains sizeable error bars, which we estimate as
$U=0.80\pm0.05$meV.

\subsection{Proximity gap}
\label{delta}
\begin{figure}[ht]
\centering
\includegraphics[width=0.8\columnwidth]{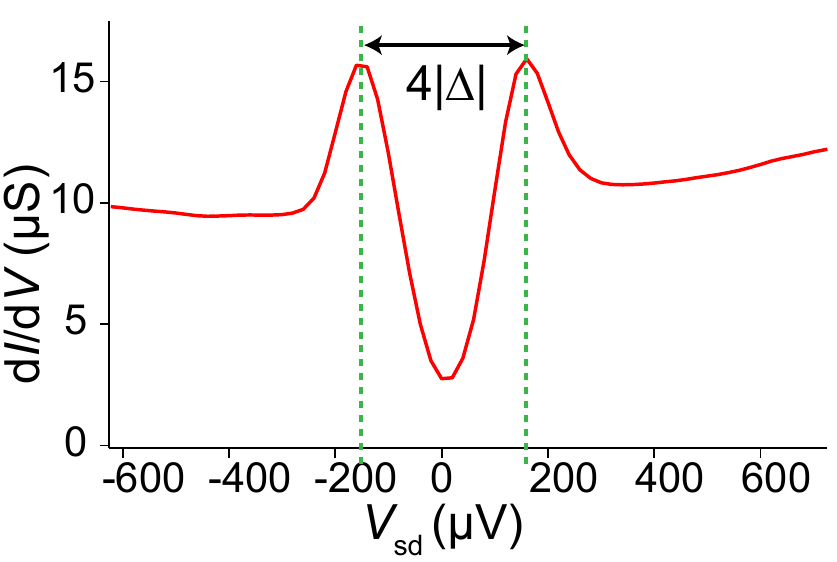}
\caption{\textbf{Signature of the superconducting $50$~nm thick aluminum
leads.} The differential conductance \textit{vs} bias voltage shows the
presence of a superconducting gap around zero-bias, as well as the
cotunneling peaks arising from the quasiparticule tunneling. The gap value 
$\Delta \approx 80~\mu$eV is thus obtained. The measurement was done in a blocked region of the nano-SQUID, which explain that no supercurrent is visible at zero bias.}
\label{fig:gap}
\end{figure}

Current-bias measurements were performed in order to directly access both the
superconducting switching current and the differential conductance
of the nano-SQUID at $T=35$mK. In the presence of superconductivity, the two 
cotunneling peaks associated with
quasiparticle current in the differential conductance~\cite{buitelaar_2002} appear at 
$V=\pm 2\Delta/e \approx\pm160\mu$V, where 2$\Delta$ is the superconducting gap 
provided by the proximity effect on the nanotube, see Fig.~\ref{fig:gap}.
This allows to extract the superconducting gap in our device, $\Delta \approx
80\mu$eV, which is reduced from the bulk value of
$\Delta_\mathrm{bulk}=175\mu$eV for aluminum, due to the
thin palladium contact layer between the carbon nanotube and the aluminum electrodes.

\section{Switching current detection}
\label{detection}

\begin{figure}[tbp]
\centering
\includegraphics[width=1.0\columnwidth]{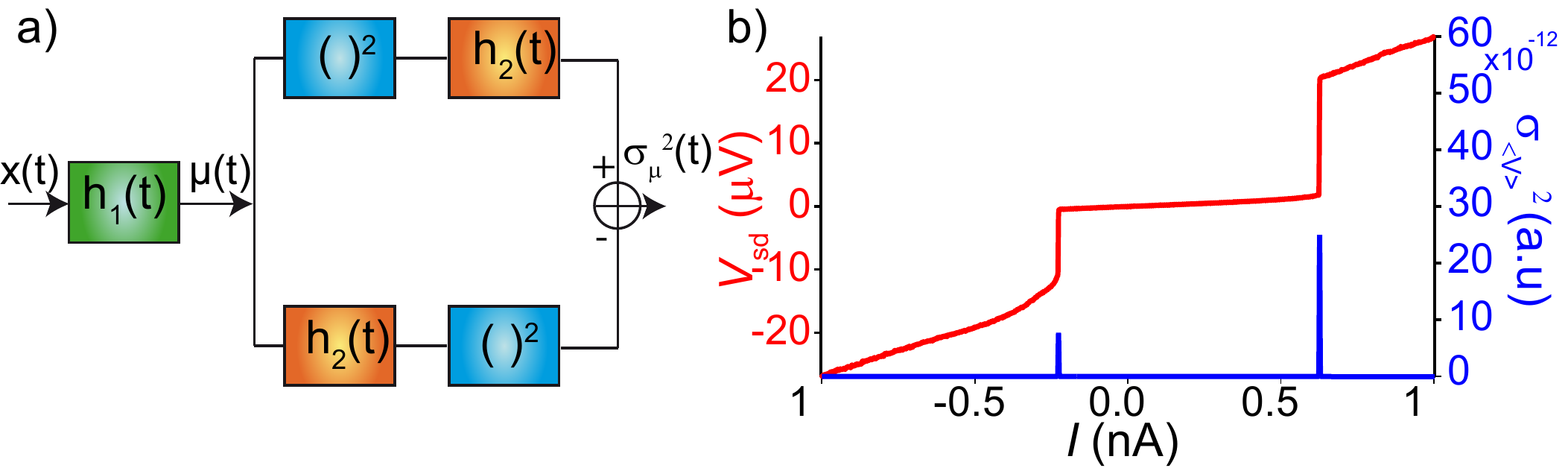}
\caption{\textbf{Digital filter for switching current detection.} 
\textbf{a)} Schematic view of the implemented filter based on the work of Liu \textit{et
al.}~\cite{liu_1995}. The first order moment $\mu (t)$ of the signal $x(t)$ in a
sliding window is obtained through the filter $h_1(t)=\mathrm{Rect}(t/L_1)/L_1$, with
$\mathrm{Rect}(t)$ the normalised rectangular function and $L_1$ the filter length. The
variance $\sigma_{\mu}^2(t)$ of $\mu (t)$ is simply calculated as $\big<\mu
(t)^2\big>$-$\big<\mu (t)\big>^2$ in a sliding window with $h_2(t)=\mathrm{Rect}(t/L_2)/L_2$. 
\textbf{b)} Voltage/Current characteristics and the estimated variance of the first 
order moment obtained by the implemented digital filter.}
\label{fig:filtre}
\end{figure}

In order to have an accurate method to extract the switching current from
voltage/current characteristics, even for small transition voltage jumps to the
dissipative state, we have implemented a digital filter based on the work of Liu
\textit{et al.}~\cite{liu_1995}. The main purpose of this filter is the
detection of transitions from noisy signal, which we apply to the
superconducting/ normal transition. The operation consists  in
estimating the variance of the first order moment of the signal in a sliding
window. A schematic view of the filter is presented in
Fig.~\ref{fig:filtre}a. The first order moment $\mu (t)$ is estimated
to begin with \textit{via} a classic averaging filter in a sliding window 
characterized by the impulse response $h_1(t)=\mathrm{Rect}(t/L_1)/L_1$ with 
$\mathrm{Rect}(t)$ the normalised rectangular function and $L_1$ the filter length.
Finally the estimated variance is obtained by $\big<\mu (t)^2\big>$ 
- $\big<\mu (t)\big>^2$ with
another averaging filter $h_2(t)$ of length $L_2$. For the switching current
detection with a sample rate of one thousand, we have taken $L_1=L_2=4$.
Such filter provides a sharp signal from the steplike features of our
voltage/current characteristics as presented n Fig.~\ref{fig:filtre}b.


\begin{thebibliography}{11}
\bibitem{franceschi_2010}
S. De Franceschi, L. Kouwenhoven, C. Sch\"{o}nenberger, and W. Wernsdorfer, \textit{Hybrid superconductor -quantum dot devices},  { Nature Nanotech.} {\bf 5}, 703 (2010).

\bibitem{jarillo-herrero_2006}
P. Jarillo-Herrero, J. A Van Dam, and L. P Kouwenhoven, \textit{Quantum supercurrent transistors in carbon nanotubes}, Nature \textbf{439}, 953–956 (2006).

\bibitem{dam_2006}
J. A. Van Dam, Y. V Nazarov, E. {P.A.M} Bakkers, S. De Franceschi, and L. P Kouwenhoven, \textit{Supercurrent reversal in quantum dots}, {Nature} {\bf 442}, 667 (2006).

\bibitem{cleuziou_2006}
{J.-P.} Cleuziou, W. Wernsdorfer, V. Bouchiat, T. Ondar\c cuhu, and M. Monthioux, \textit{Carbon nanotube superconducting quantum interference device}, {Nature Nanotech.} {\bf 1}, 53 (2006).

\bibitem{winkelmann_2009}
C. B. Winkelmann, N. Roch, W. Wernsdorfer, V. Bouchiat, and F. Balestro, \textit{Superconductivity in a 
single C60 transistor},  {Nature Physics} {\bf 5}, 876 (2009).

\bibitem{hofstetter_2009}
L. Hofstetter, S. Csonka,J. Nyg\aa rd, and C. Sch\"{o}nenberger, \textit{Cooper pair splitter realized in a two-quantum-dot Y-junction} {Nature} {\bf 461}, 960 (2009). 

\bibitem{herrmann_2009}
L. G. Herrmann, F. Portier, P. Roche, A. Levy Yeyati, T. Kontos, and C. Strunk  \textit{Carbon nanotubes as cooper-pair beam splitters}
{Phys. Rev. Lett.} {\bf 104}, 026801 (2010).



 

\bibitem{jorgensen_2007}
H. Ingerslev Jorgensen, T. Novotny, K. {Grove-Rasmussen}, K. Flensberg, and P. E Lindelof, \textit{Critical current 0-pi transition in designed josephson quantum dot junctions}, {Nano Letters} 
{\bf 7}, 2441 (2007).

\bibitem{kasumov_1999}
A. Y. Kasumov, R. Deblock, M. Kociak, B. Reulet, H. Bouchiat, I. I. Khodos, Yu. B. Gorbatov, V. T. Volkov, C. Journet, and M. Burghard, \textit{Supercurrents Through Single-Walled Carbon Nanotubes}, Science \textbf{284}, 1508-1511 (1999).


\bibitem{buitelaar_2002}
M. R. Buitelaar, T. Nussbaumer, and C. Schonenberger, \textit{Quantum dot in the Kondo regime coupled to superconductors}, Phys. Rev. Lett. \textbf{89}, 256801 (2002).


\bibitem{cleuziouCM_2006}
J.-P. Cleuziou, W. Wernsdorfer, V. Bouchiat, T. Ondar\c cuhu, 
and M. Monthioux, \textit{Tuning the Kondo effect with back and side gates - Application to carbon nanotube superconducting quantum interference devices and pi-junctions}, {Preprint} {\tt arXiv:0610622} (2006).

\bibitem{cleuziou_2007}
J.-P. Cleuziou, W. Wernsdorfer, S. Andergassen, S. Florens,V. Bouchiat, T. Ondar\c cuhu, 
and M. Monthioux, \textit{Gate-Tuned High Frequency Response of Carbon Nanotube Josephson Junctions},
{Phys. Rev. Lett.} {\bf 99}, 117001 (2007).

\bibitem{pallecchi_2008}
E. Pallecchi, M. Gaass, D. A. Ryndyk, and Ch. Strunk, \textit{Carbon nanotube Josephson junctions with Nb contacts}, {Appl. Phys. Lett.} {\bf 93}, 072501 (2008).


\bibitem{rasmussen_2009}
K. Grove-Rasmussen, H. I. J\o rgensen, B. M. Andersen, J. Paaske, T. S. Jespersen, J. Nygård, K. Flensberg, and P. E. Lindelof, \textit{Superconductivity-enhanced bias spectroscopy in carbon nanotube quantum dots}, Phys. Rev. B \textbf{79}, (2009).

\bibitem{eichler_2009}
A. Eichler, R. Deblock, M. Weiss, C. Karrasch, V. Meden, C. Schonenberger, and H. Bouchiat, \textit{Tuning the Josephson current in carbon nanotubes with the Kondo effect}, Phys. Rev. B \textbf{79}, (2009).

\bibitem {kanai_2010}
Y. Kanai, R. S. Deacon, A. Oiwa, K. Yoshida, K. Shibata, K. Hirakawa, and S. Tarucha, \textit{Electrical control of Kondo effect and superconducting transport in a side-gated InAs quantum dot Josephson junction}, {Phys. Rev. B} {\bf 82}, 54512 (2010).

\bibitem{beenakker1992}
C. W. J. Beenakker, and H. van Houten, \textit{Single-Electron Tunneling and Mesoscopic
Devices} (Springer, Berlin, 1992).


\bibitem{rozhkov1999}
A. V. Rozhkov and D. P Arovas, \textit{Josephson coupling through a magnetic impurity}, {Phys. Rev. Lett.} {\bf 82}, 2788 (1999).


\bibitem{clerk2000} 
A. A. Clerk, and V. Ambegaokar, \textit{Loss of $\pi$-junction behavior in an interacting impurity Josephson junction}, {Phys. Rev. B} {\bf 61}, 9109 (2000).

\bibitem{yoshioka_2000}
T. Yoshioka, and Y. Ohashi, \textit{Numerical renormalization group studies on single impurity Anderson model in superconductivity: a unified treatment of magnetic, nonmagnetic impurities, and resonance scattering}, {J. Phys. Soc. Japan} {\bf 69}, 1812 (2000).

\bibitem{vecino_2003}
E. Vecino, A. Martin-Rodero, and A. Levy Yeyati, \textit{Josephson current through a correlated quantum level: Andreev states and $\pi$ junction behavior}  {Phys. Rev. B} {\bf 68}, 035105 (2003).

\bibitem{siano_2004}
F. Siano, and R. Egger, \textit{Josephson current through a nanoscale magnetic quantum dot}, {Phys. Rev. Lett.} {\bf 93}, 047002 (2004).

\bibitem{choi_2004}
M. S. Choi, M. Lee, K. Kang, and W. Belzig, \textit{Kondo effect and Josephson current through a quantum dot between two superconductors}, {Phys. Rev. B} {\bf 70}, 020502 (2004).

\bibitem{sellier_2005}
G. Sellier, T. Kopp, J. Kroha, and Y. S. Barash, \textit{$\pi$-Junction behavior and Andreev bound states in Kondo quantum dots with superconducting leads}, {Phys. Rev. B} {\bf 72}, 174502 (2005).


\bibitem{novotny_2005}
Tomas Novotny, Alessandra Rossini, and Karsten Flensberg, \textit{Josephson current through a molecular transistor in a dissipative environment}, {Phys. Rev. B} {\bf 72}, 224502 (2005).

\bibitem{bauer_2007}
J. Bauer, A. Oguri, and A. C. Hewson, \textit{Spectral properties of locally correlated electrons in a Bardeen-Cooper-Schrieffer superconductor}, {J. Phys.: Condens. Matter} {\bf 19}, 486211 (2007).

\bibitem{karrasch_2008}
C. Karrasch, A. Oguri, and V. Meden, \textit{Josephson current through a single Anderson impurity coupled to BCS leads}, {Phys. Rev. B} {\bf 77}, 024517 (2008).

\bibitem{meng_2009}
T. Meng, S. Florens, and P. Simon, \textit{Self-consistent description of Andreev bound states in Josephson quantum dot devices}, {Phys. Rev. B} {\bf 79}, 224521 (2009).



\bibitem{glazman_1989}
L. I. Glazman, and K. A. Matveev, \textit{Resonant Josephson current through Kondo impurities in a tunnel barrier}, {JETP Lett.} {\bf 49}, 659 (1989).

\bibitem{kouwenhoven_2001}
L. Kouwenhoven, and L. Glazman, \textit{Revival of the Kondo effect}, {Phys. World} {\bf 14}, 33 (2001).



\bibitem{liu_1995}
Liu, W. Y., Magnin, I. E. \& Gimenez, G. \textit{A New Operator for the Detection of Transitions in Noisy Signals}, {Traitement du Signal} {\bf 12}, 225 (1995).



\bibitem{grabert_1992}
H. Grabert, and M. H. Devoret, \textit{Single Charge Tunneling}, {Plenum press, New York, 1992}.


\bibitem{haldane_1978} 
F. D. M. Haldane, \textit{Scaling Theory of the Asymmetric Anderson Model}, {Phys. Rev. Lett.} {\bf 40}, 416 (1978).


\bibitem{bulla2008}
R. Bulla, T. A. C. Costi, and T. Pruschke, \textit{Numerical renormalization group method for quantum impurity systems}, {Rev. Mod. Phys.} {\bf 80}, 395 (2008).

\bibitem{fujii_2003}
T. Fujii, and K. Ueda, \textit{Perturbative approach to the nonequilibrium Kondo effect in a quantum dot}, {Phys. Rev. B} {\bf 68}, 1 (2003).

\bibitem{goldhaber_1998}
D. Goldhaber-Gordon, J. Göres, M. Kastner, Hadas Shtrikman, D. Mahalu, and U. Meirav, \textit{From the Kondo Regime to the Mixed-Valence Regime in a Single-Electron Transistor}, { Phys. Rev. Lett.} {\bf 81}, 5225 (1998).



\bibitem{rocca_2007}
M. Della Rocca, M. Chauvin, B. Huard, H. Pothier, D. Esteve, and C. Urbina, \textit{Measurement of the Current-Phase Relation of Superconducting Atomic Contacts}, {Phys. Rev. Lett.} {\bf 99}, 127005 (2007).


\bibitem{pillet_2010}
J.-D. Pillet, C. H. L. Quay, P. Morfin, C. Bena, A. Levy Yeyati, and P. Joyez, \textit{Andreev bound states in supercurrent-carrying carbon nanotubes revealed}, {Nature Physics} {\bf 6}, 965 (2010).



\end{thebibliography}
\end{document}